\documentclass[11pt,preprint]{aastex}
\usepackage{epsfig}
\usepackage{natbib}
\newcommand{\htwo}{H$_2$}

\newcommand{\eg}{{\it e.g.,}}
\newcommand{\cf}{c.f.}

\newcommand{\kms}{km~s$^{-1}$}
\newcommand{\cc}{cm$^{-3}$}

\newcommand{\hst}{\textit{HST\/}}
\newcommand{\fuse}{\textit{FUSE\/}}

\newcommand{\te}{$T_{e}$}
\newcommand{\ti}{$T_{i}$}
\newcommand{\nelec}{$n_{e}$}
\newcommand{\rj}{$R_{\rm J}$}

\shorttitle{The Far Ultraviolet Spectrum of the Io Plasma Torus}
\shortauthors{Feldman et al.}

\begin{document}

\title{The Far Ultraviolet Spectrum of the Io Plasma Torus}  

% USE FULL NAME
\author{Paul D. Feldman, Darrell F. Strobel\altaffilmark{1}, H. Warren
Moos, and Harold A. Weaver\altaffilmark{2}}

\affil{Department of Physics and Astronomy, The Johns Hopkins University\\ 
Charles and 34th Streets, Baltimore, MD 21218}
\email{pdf@pha.jhu.edu}

\altaffiltext{1}{and Department of Earth and Planetary Sciences, The Johns
Hopkins University}

\altaffiltext{2}{present address: Space Department,
Johns Hopkins University Applied Physics \mbox{Laboratory,}
11100 Johns Hopkins Road,
Laurel, MD 20723-6099}

%\pagestyle{myheadings}
%\markright{ \today}

%%%% SEE BELOW FOR HOW TO ADD AUTHORS WITH ALT AFFILIATIONS

\begin{abstract}

The spectrum of the Io plasma torus in the range 905--1187 \AA\ was
recorded at 0.26~\AA\ resolution by the {\it Far Ultraviolet Spectroscopic
Explorer} (\fuse) on 2001 January 14.  Five orbits of data were obtained
with the west ansa of the torus centered and tracked in the $30''
\times 30''$ apertures of the FUSE spectrographs for a total
observation time of 9740 seconds.  This region of the spectrum is
dominated by transitions of \ion{S}{2}, \ion{S}{3} and \ion{S}{4},
whose multiplet structure is nearly completely resolved.  We confirm
our earlier detection of emission from resonance multiplets of
\ion{Cl}{3} and \ion{Cl}{2} and derive an abundance of Cl$^{+2}$ of
2.1\% relative to S$^{2+}$, leading to an overall chlorine ion
abundance in the torus of slightly less than 1\%.  A number of features
near 990 \AA\ remain unidentified, and \ion{C}{3} $\lambda$977 is
detected in two independent channels at the 3-$\sigma$ level.  The
inferred relative ion abundance of C$^{2+}$ relative to S$^{2+}$ is
$3.7 \times 10^{-4}$.  We also present spectra at 0.085~\AA\ resolution
taken on 2001 October 19 and 21 with the $4'' \times 20''$ aperture.
In these spectra the observed lines are resolved and their widths
correspond to ion temperatures of 60--70~eV for all three sulfur ions.

\end{abstract}

%\keywords{topics alphabetically --- subsubtopic: topic --- subtopic, subtopic}

\newpage
\section{INTRODUCTION}

In a previous {\it Letter} \citep[][hereinafter Paper I]{Feldman:2001a}
we reported on early spectroscopic observations of the Io plasma torus
with the {\it Far Ultraviolet Spectroscopic Explorer} (\fuse).  Those
observations were made in January 2000 before \fuse\ acquired the
capability to track moving solar system targets yet yielded new
information about the content of the torus and in particular allowed
the determination of the relative abundance of chlorine ions from the
first detection of the ultraviolet resonance transitions of both
\ion{Cl}{2} and \ion{Cl}{3}.  
The source of chlorine on Io is currently of interest following the
detection of NaCl in Io's atmosphere \citep{Lellouch:2003} and recent
detailed modelling of volcanic chemistry \citep{Moses:2002}.
As part of the \fuse\ team solar system
project, the torus was observed again, this time with full tracking
capability, on 2001 January 14, during a Jovian campaign timed to
coincide with the near closest approach of {\it Cassini} to Jupiter.
The spectra obtained were a significant improvement over the earlier
data.  A clear enhancement of signal-to-noise ratio by a factor of four
enabled confirmation of the identification of the weak \ion{Cl}{2}
emission reported earlier, the detection of many weak \ion{S}{2} and
\ion{S}{3} emissions including several from very highly excited states,
and the spectrum between 905 and 990 \AA\ that was not obtained from the
earlier data due to thermal misalignment of the SiC channels.
Additional, higher resolution spectra of the strongest emissions were
obtained during two series of limb scans of Jupiter in October 2001.
 
In this paper we present a more detailed analysis of the spectrum and
in particular examine the torus abundances that can be inferred from
the presence or absence of emissions of ions of C, N, Si and P, all of
which have strong transitions in the \fuse\ spectral range.  A number
of weak spectral features remain unidentified and there are a few cases
where the spectral assignment of the observed lines are uncertain.  The
higher resolution spectra show line widths that are greater than the
instrumental widths which allows us to directly determine the ion
temperature in the warm region of the torus where the ultraviolet
emissions are produced.  A comparison with the earlier data shows a
significant change in relative ionization stages of sulfur in the year
between observations.  Temporal variations on scales of days to months
have been reported from extended {\it Cassini} UVIS observations made
around the time of Jupiter fly-by \citep{Steffl:2003a}.  The abundance
of chlorine ions relative to those of sulfur also appears to have
changed.

\section{OBSERVATIONS}

Observations of the west ansa of the torus were made on 2001 January
14, beginning at 05:56:40 UT.  A total exposure of 9740 seconds was
obtained by accumulating spectra over 5 contiguous orbits.  The
center of the $30'' \times 30''$ aperture (LWRS) was placed at the position of
the ansa at the beginning of each exposure and both the motion of
Jupiter on the sky and the wobble of the torus were tracked during each
individual exposure.  Tracking was verified from the constancy of the
count rates in the principal ion emissions over each $\sim$2000 s exposure
derived from the time-tagged data files.  There is no spatial
information about the distribution of the emission within the
aperture.

At the time of the observation, Jupiter's heliocentric distance was
5.05 AU and its geocentric distance was 4.40 AU.  Details of the \fuse\
instrumentation have been given by \citet{Moos:2000} and
\citet{Sahnow:2000}.  Basically, \fuse\ consists of four separate
telescope/spectrograph assemblies, two employing lithium fluoride
coatings (LiF) and two using silicon carbide (SiC).  Each of four
separate detectors (denoted 1a, 1b, 2a, and 2b) records two spectra,
one each from a LiF and a SiC channel spanning a given wavelength
interval, giving a total of eight separate spectra.
Due to the extended but finite size of the torus
emitting region within the aperture, the effective spectral resolution
is 0.26 \AA.  The data presented here were reprocessed with the \fuse\
pipeline, calfuse version 2.2.1, all five orbits of data were co-added
together, and the extracted fluxes were converted to average brightness
(in rayleighs) in the $30'' \times 30''$ aperture.  The wavelengths
were reduced to laboratory values by first taking out a heliocentric
velocity correction applied by the pipeline and then removing the
doppler shifts due to the Earth's velocity relative to Jupiter and
the plasma torus co-rotation of 75 \kms.  The agreement with laboratory
wavelengths in all cases is better than 0.05 \AA, and testifies to the
accuracy of the wavelength calibration achieved in this version
of the pipeline.

Drift scans of the eastern limb of Jupiter were made on 2001 October 19
and 21 using the \fuse\ MDRS ($4'' \times 20''$) aperture.  A single
scan was made over each of five consecutive $\sim$3000 s orbits on both
days.  Each scan began with the slit, with the long dimension aligned
nearly north-south, placed on a chord centered on the Jovian equator
near the eastern limb.  The spacecraft then executed a slew so that the
apparent motion of the slit on the sky was towards the east,
perpendicular to the long dimension of  the slit, at a rate of
$\sim$0.23 arc-seconds per minute.  At the end of each scan the slit
was completely off the disk of the planet.  For these observations,
Jupiter's heliocentric distance was 5.14 AU and its geocentric distance
was 4.86 and 4.83 AU, respectively, on the two dates.

The goal of these observations was to record the spectrum of the disk,
primarily resonance scattering by atomic hydrogen and electron impact
of \htwo.  This approach was chosen to minimize the possibility of
contamination by the much stronger polar aurorae and to take advantage
of possible limb brightening of the \ion{H}{1} emissions.  In this
respect, the foreground torus emissions were contaminants as they had
been in the lower resolution ($\sim$3 \AA) observations of Jovian
airglow by the Hopkins Ultraviolet Telescope \citep{Feldman:1993}.  At
the higher resolution of \fuse\ they are easy to separate from the disk
emissions.  

Due to thermal misalignments between channels only the the LiF1 channel
appears to show the expected level of disk emission.  This is the channel
that is coupled to the Fine Error Sensor used for guiding and
tracking.  The SiC1 channel also shows disk emission but at considerably
lower signal/noise as the effective area of this channel is a factor of
4 lower than that for LiF1.  The LiF2 and SiC2 channels both appear to
have been off the disk for the entire observation, showing only torus
emissions.  For the LiF1 channel the time tagged data clearly show the
disk emissions decreasing with time while the torus emissions increase
as more and more of the background emission becomes visible in the
aperture.  It is also possible to follow the brightness variation of
the torus emissions over successive orbits as the centrifugal equator
of the torus wobbles through Jupiter's equatorial plane at the limb.
Because the emissions are about a factor of two weaker than at the
ansae of the torus and because of the lower throughput of the MDRS
aperture the torus spectra presented below are derived by co-adding all
of the five orbits on each day for a total integration time of 31,333
seconds.

\section{SPECTRA}

\subsection{LWRS Spectra}

An overview spectrum from 995 to 1105 \AA\ obtained from the January
2000 observations was given in Paper~I.  The spectra from January 2001
are essentially similar except for the factor of four increase in
signal-to-noise ratio.  This is particularly important for confirming
the \ion{Cl}{2} and \ion{Cl}{3} multiplets which are shown in
Figure~\ref{spec1}.  The chlorine emissions, both of which are
principal resonance transitions, occur close to the wavelengths of
resonance transitions of \ion{S}{3} and \ion{S}{4}.  The components of
the \ion{Cl}{3} triplet appear to be in the ratio 1:2:3 expected for
the optically thin branching ratios even though the brightness of the
1015 \AA\ line is somewhat uncertain due to blending with the strong
\ion{S}{3} lines nearby.  Similarly, five of the six lines of the
\ion{Cl}{2} multiplet are clearly identified, again with the expected
branching ratios.  A list of all of the identified sulfur and chlorine
ion emissions is given in Tables~\ref{tab:sii} and \ref{tab:siii}.
Note that in two instances the atomic configurations are either not
given or are uncertain.

\placefigure{spec1}
\placetable{tab:sii}
\placetable{tab:siii}

The tables also give the identifications and brightnesses of a number
of lines not identified in the January 2000 spectra
(Paper~I), particularly of \ion{S}{2}.  This is not only due
to the enhanced S/N but also to the absence of thermal misalignment
between the LiF and SiC channels that occurred in the earlier
observation.  A segment of the SiC1b spectrum, which extends to the
\fuse\ short wavelength limit at 905 \AA\ and includes the second
resonance multiplet of \ion{S}{2}, is shown in Figure~\ref{spec2}.
Figure~\ref{spec3} shows a region of the spectrum from the LiF2a
channel containing highly excited multiplets of \ion{S}{3} and
\ion{S}{4} that are extremely useful as temperature diagnostics (see
Section~\ref{diag}).
 
\placefigure{spec2}
\placefigure{spec3}

\subsubsection{Unidentified Features}

The high sensitivity and spectral resolution of this data set make possible the
search for and identification of many other ionic species that have
resonance transitions in the \fuse\ spectral range.  A first
examination of the data indicates that almost all of the observed
emissions can be identified with known and published transitions of
sulfur and chlorine.  There are a few lines whose assignment is
uncertain such as the \ion{S}{3} line at 1016.85 \AA\ seen in the top
panel of Figure~\ref{spec1} and listed in Table~\ref{tab:siii}.
\citet{Kelly:1987} gives a wavelength that matches the observed line
very well but does not assign the transition.  A weaker line seen in
the same figure at $\sim$1012.0 \AA\ is also listed by
\citeauthor{Kelly:1987} as a \ion{S}{3} line at 1011.91 \AA, again
without an assignment.  The other ambiguity is the line at 1088.8
\AA\ identified by \citeauthor{Kelly:1987} as \ion{S}{4}.  However, as
we noted in our earlier paper, \citeauthor{Kelly:1987} assigns this
line to the same transition as the 1098.9 and 1099.5 \AA\ lines of
\ion{S}{4} given in the NIST Atomic Spectra
Database\footnote{http://physics.nist.gov/}.  Of these latter two
lines, Kelly lists only the 1099.5 \AA\ line, and does not identify the
transition, so his assignment of the 1088.8 \AA\ line is suspect.  The
NIST Database does list a \ion{S}{4} transition at 1088.214 \AA\
between high-lying $^4{\rm D}^{\rm o}$ and $^4{\rm F}$ states, the
upper one requiring $\sim$39 eV of excitation energy.  However, the
wavelength differs by 0.6 \AA\ from our measured wavelength and a
companion line at 1091.18 \AA\ is absent, so this identification is
unlikely.

The only truly unidentified features in the spectrum occur near 990
\AA.  Figure~\ref{spec4} shows the spectral region from 970 to 1000
\AA\ from three separate channels including the very short wavelength
range of the LiF1a channel.  Five distinct lines are seen in the SiC2a
spectrum just shortward of the \ion{S}{2} line at 996.0 \AA, and the
other spectra, where the wavelengths overlap, show the same features
although at poorer S/N.  These lines are at 991.5, 992.9, 993.5, 994.5,
and 995.1~\AA\ and have brightnesses in the range of 0.3--0.6
rayleighs.  It is tempting to assign the feature at 991.5 \AA\ to
\ion{N}{3} $\lambda$991.56 but this is questionable due to the absence
of the companion \ion{N}{3} line at 989.79 \AA\ which should be of
comparable brightness.  \citeauthor{Kelly:1987} lists an unassigned
\ion{S}{2} transition at 991.73~\AA, but the wavelength agreement is
not quite as good.  No obvious single candidate multiplet for all of
the lines has been found in the available spectroscopic databases.  Of
the other sulfur ions, there is a high-lying transition of \ion{S}{5}
at 994.46 \AA\ but this is considered unlikely since it requires 56 eV
excitation energy and S$^{4+}$ is present only at about 2\% relative to
S$^{2+}$ \citep{Steffl:2003a}.  There is a \ion{Si}{3} triplet with
lines at 993.5 and 994.8~\AA, close to the wavelengths of two observed
lines, but the strongest member of the triplet should be at 997.4 \AA,
which is missing.  There should also be a stronger \ion{Si}{3}
multiplet at 1108--1113~\AA.

\placefigure{spec4}

\subsubsection{Other Expected Features}
\label{other}

Both singly- and doubly-ionized carbon have strong transitions in the
\fuse\ spectral range and these were searched for in the January 2001
data.  The \ion{C}{3} line at 977.02 \AA\ appears to be present
in both SiC channel spectra (Figure~\ref{spec4}) at the 3-$\sigma$
level.  Combining the two spectra gives a brightness of $0.19 \pm
0.034$ rayleighs.  The \ion{C}{2} doublet at 1036.34 and 1037.02 \AA\
is not detected with a 1-$\sigma$ upper limit of 0.04 rayleighs.
Similar limits apply to the \ion{Si}{3} multiplet, the strongest
members of which would be at 1108.36, 1109.97, and 1113.23 \AA.
Phosphorus, which has about the same relative solar abundance as chlorine
\citep{Grevesse:1998}, also has strong transitions of its low
ionization states, \ion{P}{2} at 1154 \AA, \ion{P}{3} at 918 and 1002
\AA, and \ion{P}{4} at 951 \AA.  However, none of these multiplets are
detected.  The possible identification of an \ion{N}{3} feature was
mentioned above, but also likely to be present is \ion{N}{2}, whose
resonance multiplet is at 1085 \AA, but which is also not detected.

The resonance transition of \ion{S}{5} at 786 \AA\ has recently been 
reported by \citet{Steffl:2003a}.
There are only high-lying transitions of \ion{S}{5} in the \fuse\
spectral range.  An intercombination transition that connects to the
ground state lies just beyond the long wavelength limit of the LiF1b
detector at 1188 \AA.  While there are also no transitions of either
\ion{O}{2} or \ion{O}{3} connecting to their ground states there are
several transitions between excited states of \ion{O}{2} that could be
observed.  One such triplet, with its strongest component at 1083.13
\AA, has been identified in \fuse\ spectra of terrestrial airglow where
its source is likely photoionization of neutral atomic oxygen
\citep{Feldman:2001b}.  The upper state of this transition lies 26.3 eV
above the ground state and might be expected as we do observe
\ion{S}{3} $\lambda$1126 and \ion{S}{4} $\lambda$1099, which require
21.5 and 23.0 eV, respectively, for excitation (see
Figure~\ref{spec2}).  The \ion{O}{2} multiplet, however, like the
\ion{N}{2} resonance multiplet, falls in the gap between the two
segments of each LiF detector and can be observed only with the SiC
channels, whose effective areas at these wavelengths are a factor of
five or more lower.  There are also no collision strengths available
for this \ion{O}{2} transition.

\subsection{MDRS Spectra}
\label{mdrs}

A portion of the higher resolution spectra corresponding to the top
panel of Figure~\ref{spec1} is shown in Figure~\ref{spec_mdrs}.  The
lower spectrum is from LiF1a and contains some weak disk emission
features in addition to the torus lines.  The upper spectrum is from
LiF2b and shows only torus lines but is considerably noisier due to the
factor of two lower effective area of this channel relative to LiF1a.
In order to obtain the best S/N possible, all of the data from both
dates of observation have been co-added for each of the channels.
While the LiF2b spectra are useful in identifying torus features, only
the LiF1a spectra are used for fitting the line profiles.

\placefigure{spec_mdrs}

Of the six lines in the \ion{S}{3} multiplet only the lines at 1015.496
and 1015.561 \AA\ are not resolved.  The widths of the lines for this
multiplet, the \ion{S}{4} multiplet at 1062--1073 \AA\ and the 
\ion{S}{2} line at 1045.76 \AA\ are fit to gaussians and give linewidths
of 0.14--0.15 \AA.  The intrinsic linewidth for emission uniformly
filling the MDRS aperture is evaluated from the terrestrial airglow
triplet of \ion{O}{1} at 1040 \AA, and found to be 0.085 \AA.  The
observed widths then correspond to the convolution of a gaussian of $7
- 8 \times 10^5$ K with the instrumental width for all of the three
sulfur ions.  This translates to an ion temperature of 60--70 eV.

\section{DISCUSSION}

\subsection{Comparison with January 2000 Data}
\label{comp}

In Paper~I the spectrum was fit to an isothermal multi-level emission
model using version 3 of the CHIANTI database and software
\citep{Dere:1997,Dere:2001} that was updated with recent calculations
of sulfur ion collision strengths that are now included in version 4
\citep{Young:2003}.  Normalizing to \ion{S}{3}, the relative ion
emissions were fit with an electron temperature, $T_e$, of 80,000~K
(6.9~eV), and relative abundances of
S$^{+}$\,:\,S$^{2+}$\,:\,S$^{3+}$ = 0.30\,:\,1.00\,:\,0.23.  The
present spectrum, including the spectral region below 995 \AA\ not
observed in the January 2000 data, is again well fit with $T_e =
80,000$~K, but with different relative abundances,
0.41\,:\,1.00\,:\,0.17.  Comparing the ratios of the strongest lines in
each observation, we find that relative to S$^{2+}$, S$^{+}$ has
increased by 37\% while S$^{3+}$ has decreased by 26\%.  Further,
comparing the relative ion intensities in the MDRS spectra from October
2001, we find that the relative abundances are about the same as in
January 2001, though this does not imply that they were stationary over
the ten month interval separating the observations.  

Our results may be compared directly with those derived from a series
of observations of the torus made by the {\it Cassini} ultraviolet
spectrometer on 2001 January 6, 12, and 14, recently reported by
\citet{Steffl:2003b}.  From a composite spectrum from all three days
spanning the wavelength interval from 650 to 1750~\AA, and using the
current version of the CHIANTI database, they find an electron
temperature of 4.6~eV and relative abundances of
S$^{+}$\,:\,S$^{2+}$\,:\,S$^{3+}$ = 0.30\,:\,1.00\,:\,0.15 at
6.3~\rj.  They also simulate the suprathermal electron tail
($\kappa$-distribution) by a superposition of three Maxwellians but
find little difference in the derived abundances.  As noted below in
Section~\ref{diag}, the lower value of \te\ found by
\citeauthor{Steffl:2003b} is not surprising and is in fact in good
agreement with the result from the reanalysis of the HUT data, also
spanning a wider wavelength range than the \fuse\ spectrum, given in
Paper~I.  The difference in derived relative S$^{+}$ abundance is likely
due to the larger field-of-view of the \fuse\ aperture as \citeauthor{Steffl:2003b} show its abundance to be steadily increasing
inwards of 6.3~\rj.

\subsection{Electron Density and Temperature Diagnostics}
\label{diag}

The far ultraviolet emissions observed by \fuse\ originate in the
``warm torus'', the region outward from $\sim$5.8 \rj, where the plasma
is characterized by an electron density, \nelec, $\sim$2000 \cc, an
electron temperature, \te, $\sim$5 eV (1 eV = $1.16 \times 10^4$ K),
and an ion temperature, \ti, in the range of 50--100 eV
\citep[\cf][]{Bagenal:1994}.  These parameters vary significantly with
radial distance from Jupiter and also with height above the centrifugal
plane as determined from both {\it in situ} measurements and remote
long-slit spectroscopy, the latter primarily from ground-based
observations of \ion{S}{2} and \ion{S}{3} \citep[\eg][]{Thomas:2001},
which sample all of the torus.  The remote observations exploit plasma
diagnostics in the form of temperature or electron density sensitive
spectral line ratios \citep{Hall:1994b} and are independent of
inferences based on the total flux in a given emission feature
\citep{Taylor:1995}.  As noted above in Section~\ref{comp}, we use
the CHIANTI database for the atomic data, primarily transition
probabilities and collision strengths, for the evaluation of the
requisite line ratios. 

There are, however, several caveats in using these diagnostics with the
\fuse\ spectra.  The relatively large field-of-view, $30'' \times
30''$, translates into a range of radial distance and height above the
centrifugal plane of $\pm 0.67$ \rj\ over which these parameters are
averaged as there is no spatial resolution within the aperture.  This
is in addition to the line-of-sight integration that is common to all
remote observations.  On the other hand, the derived parameters well
characterize the regions where the emission is being produced.  

Another problem, which has been pointed out before, is that the use of
theoretical, thermally averaged effective collision strengths assumes a
Maxwellian electron velocity distribution whereas the torus plasma is
known to have a suprathermal tail that could contribute significantly
to the excitation of the ultraviolet emissions.
\citet{Sittler:1987} found small departures from a Maxwellian
distribution at Io's orbital distance with a suprathermal (\te\
$\sim$1~keV) to thermal (\te\ $\sim$5~eV) electron density ratio of
$\sim$0.001.  
Further support for non-thermal electrons comes from the {\it Galileo}
plasma science instrument \citep{Frank:2000}.  They found thermal core
temperatures varying from 3 eV at $\sim$6 \rj\ to 8 eV at 7.8 \rj, with
departures from the core Maxwellian distribution for energies in excess
of 30 eV.
In an analysis of {\it Ulysses} spacecraft data for the hot
torus at 7--9 \rj, \citet{Meyer-Vernet:1995} applied the theory of
velocity filtration for non-Maxwellian plasmas given by
\citet{Scudder:1992}, to the centrifugal force potential constraining
torus plasma.  They used a generalized Lorentzian function, known as
the ``kappa'' distribution, to represent a non-thermal distribution
function that is quasi-Maxwellian at low energies but has a power-law
decrease (with the exponent a function of $\kappa$) at high energies.
This distribution has been used to explain various features of both
{\it Voyager} data \citep{Moncuquet:2002} and \hst\ observations of
Io's ultraviolet limb glow \citep{Retherford:2003}.

Thus, there is significant evidence in both {\it in situ} and remote 
sensing observations for non-Maxwellian plasma distributions in the Io 
torus in addition to the fact that \ti\ $\gg$ \te.  
With the small $\kappa$ indices $\sim$2--3.5 inferred by
\citet{Meyer-Vernet:1995}, \citet{Moncuquet:2002}, and
\citet{Retherford:2003}, the electron temperature increases from its
value at the centrifugal equator, for example, 5 eV to $\sim$7.5 and 15
eV at 1 plasma scale height above and below the equator for $\kappa$ =
3.5 and 2, respectively.  With the large aperture of \fuse, the derived
effective electron temperature will depart significantly from its
equatorial value.

This effect is less important for the longer wavelength ultraviolet
emissions, and evidence for this effect may be found in Paper~I in the
reanalysis of the torus spectra taken by the Hopkins Ultraviolet
Telescope (HUT) during the {\it Astro-1} and {\it -2} missions
\citep{Moos:1991} using the current set of atomic data.  For both of
these data sets, the best fit to the emissions in the \fuse\ spectral
range together with \ion{S}{2} $\lambda$1256, \ion{S}{3}]
$\lambda$1729, and \ion{S}{4} $\lambda$1406, gave a value of \te\ of
$\sim$50,000 K, similar to what was found earlier by
\citet{Hall:1994b}.  Nevertheless, the derivation of an ``equivalent''
electron temperature is extremely useful for comparison with
theoretical models of the electron energy distribution and with {\it in
situ} measurements from {\it Galileo}.

A final caveat rests with the atomic data, the uncertainties not being
immediately quantifiable.  There are a number of multiplets where the
relative intensities within the multiplet as measured differ from the
model synthetic spectrum and these lines are neither \nelec\ or \te\
dependent.  There are also deviations from the predicted multiplet
relative intensities, particularly for \ion{S}{2}, but also for 
\ion{S}{3} $\lambda$1077.2 relative to the resonance multiplet.
These deviations lie mainly in the details and may serve the purpose of
being an ``astrophysical laboratory'' for fine-tuning the
calculations.

The strongest electron temperature dependence occurs for the multiplets
of \ion{S}{3} and \ion{S}{4} shown in Figure~\ref{spec2}.  As noted in
Section~\ref{other}, these multiplets require 21.5 and 23.0 eV,
respectively, for excitation, while the resonance multiplets require
only $\sim$11 eV.  Figure~\ref{tedep} shows the calculated
\te\ dependence of three lines of \ion{S}{3} and two lines of
\ion{S}{4}, together with the observed ratios from the LiF1b (crosses)
and LiF2a (diamonds) channels.  For all but one of these lines the two
channels give nearly identical results.  The discrepancy in the
\ion{S}{3} $\lambda$1126.8 blend may be the result of a known
calibration problem with the LiF1b channel \citep{Sahnow:2000}.
The data are fit by a \te\ ranging from 65,000 to 90,000 K.  However,
note that lines originating in the same multiplet give different values
of \te, the difference well exceeding the statistical uncertainties in
the measured line ratios.  This limited set of data also does not
provide any evidence that the \ion{S}{4} emissions are produced in a
region of higher \te\ than those of \ion{S}{3}, as \citet{Hall:1994b}
had concluded from the HUT data.  A complete reanalysis of the HUT
spectra will be presented elsewhere.

\placefigure{tedep}

There are a limited number of electron density diagnostic intensity ratios
available that are relatively insensitive to \te.  Figure~\ref{nedep}
shows two for \ion{S}{2} and \ion{S}{3}, calculated with version 4 of
CHIANTI for \te\ = 50,000 and 80,000 K, together with the observed
ratios.  All but one are insensitive to \te.  The two \ion{S}{3} ratios
give widely differing values for \nelec, and all give densities in the
range of 3000 -- 5000 \cc, higher than the {\it Voyager}-based
model value of 2000 \cc, but consistent with {\it in situ Galileo} PLS
measurements of of $3600 \pm 400$ \cc\ 
\citep{Frank:1996,Bagenal:1997}.  The small slopes of the ratio curves
make these calculated ratios extremely sensitive to small uncertainties
in the atomic parameters and in fact there are significant differences
in some of these curves between versions 3 and 4 of CHIANTI.
 
\placefigure{nedep}

\subsection{Ion Temperatures}

The spectra presented above in Section~\ref{mdrs} are the first
ultraviolet spectra of the Io plasma torus with sufficient spectral
resolution to allow the determination of ion temperatures from the
measured line shapes.  \citet{Thomas:2001} have used ground-based high
dispersion observations of \ion{S}{2} and \ion{S}{3} in the red and
near infrared to determine \ti\ as a function of radial distance from
Jupiter.  Unlike those measurements, the spectra described in
Section~\ref{mdrs} were obtained with the \fuse\ long-slit at
$\sim$1~\rj, oriented roughly normal to the planet's equatorial plane.
This results in a factor of two loss in flux over placing the slit at
the ansa \citep{Sandel:1982,Steffl:2003a} but effectively integrates
the emission over all radial distances.  The derived ion temperature
thus refers to the ions in their region of maximum emission, roughly
6.0--6.5 \rj.

We found that the linewidths, which are determined mostly by the
distribution of gyrovelocities, corresponded to perpendicular ion
temperatures of 60--70~eV for all three sulfur ions.
This is in excellent agreement
with the values of \ti\ for \ion{S}{2} and \ion{S}{3} in the region
from 6.0 to 6.5 \rj\ found by \citet{Thomas:2001}.  However, it is in
strong disagreement with the results from the {\it Galileo} plasma
spectrometer which \citet{Crary:1998} interpreted to give
\ti\ ~=~10--24 eV over this range of radial distance.

In addition to effects of non-Maxwellian plasma distributions 
along the magnetic field, the processes that add ion mass and energy to
the torus plasma, ionization of neutrals that escape Io by electron
impact and charge exchange, create a highly non-Maxwellian distribution
of gyro-velocity.  Fresh ions are accelerated by Jupiter's co-rotational
electric field from essentially Io's orbital velocity to co-rotation
with Jupiter and acquire a gyro-velocity equal to the velocity
difference.  Because most ions are created at Io's orbital distance,
the gyro-velocity distribution functions for newly created ions is
approximately a delta function with only a narrow spread in velocity.
\citet{Smith:1985} demonstrated in a Fokker-Planck calculation of
distribution functions for perpendicular (to the magnetic field)
velocity that if the residence time of ions in the torus is short
compared with the time to collisionally evolve into Maxwellian
distributions, the ion distribution functions remain highly
non-Maxwellian.  For conditions at the time of {\it Voyager-1}, they
found that \ion{S}{2} and \ion{O}{2} departed significantly from
Maxwellian distributions and that the average energies/temperatures
were highest for \ion{O}{2} and \ion{S}{2} and coldest for \ion{O}{3}
and \ion{S}{4} with differences approaching a factor of 2.  The latter
ions had essentially Maxwellian distributions.  \citeauthor{Smith:1985}
fit Maxwellian distributions to the quasi-thermal cores of their
calculated distribution functions and found common core temperatures
for sulfur ions under Voyager 1 conditions of 60--74 eV.  Only
\ion{O}{2} yielded a high core temperature of 105 eV.

Our procedure of fitting gaussians to the observed line widths captures
the quasi-thermal cores of the sulfur ion distribution functions.  
The data have insufficient S/N to permit detection of the non-thermal
component, which in the {\it Voyager-1} case, according to
\citeauthor{Smith:1985}, 5\% of \ion{S}{2} were in a suprathermal tail
with a distinct pickup ion signature.  The ion temperatures of
60--70~eV inferred from the \fuse\ spectra are comparable to the {\it
Voyager-1} encounter ion temperatures derived by \citet{Bagenal:1994}
from {\it Voyager} PLS data and with the theory of \citet{Smith:1985}.

\subsection{Relative Minor Ion Abundances}

As in Paper~I, we can estimate the relative abundances of minor
ions in the torus by comparing the excitation rates of these ion lines
to that of the \ion{S}{3} line at 1012.5 \AA.  Repeating from that
paper, the collisional excitation rate is given by:
%\begin{equation}
\[ C_i = \frac{const}{T_e^{1/2}}\Upsilon(T_e)e^{-E_i/kT_e} \]
%\end{equation}
\noindent 
where $\Upsilon(T_e)$ is the electron temperature dependent effective collision
strength, $E_i$ is the energy of the upper state of the transition, and
$k$ is Boltzmann's constant.  Values of $\Upsilon(T_e)$ are taken
largely from the CHIANTI 4.0 database \citep{Young:2003} except for the
recent calculations for \ion{Cl}{3} given by \citet{Ramsbottom:2001}
which is not in the current version of the database.  These include the
calculations of \citet{Tayal:1999b} for \ion{S}{3}, \citet{Blum:1992}
for \ion{C}{2}, \citet{Berrington:1989} for \ion{C}{3} and
\citet{Dufton:1989} for \ion{Si}{3}.  We find that
[Cl$^{2+}$]/[S$^{2+}$] = $0.021 \pm 0.0035$, or [Cl$^{2+}$]/(all sulfur
ions) = $0.013 \pm 0.0022$.  These are somewhat lower than the values
of 0.031 and 0.020, respectively, given in Paper~I.  We still lack the
atomic data for \ion{Cl}{2} to definitively show that Cl$^{2+}$ is the
principal ionization stage of chlorine in the torus.  Note that our
derived abundance is comparable to the cosmic abundance ratio of
[Cl]/[S] = 0.015 given by \citet{Grevesse:1998}.  As before, since
sulfur can account for no more than half of the total ion density in
the torus this leads to the conclusion that chlorine ions are present
at the level of 1\% or less of the total ion population.  There
appears to be a statistically significant decrease of $\sim$30\% in the
Cl$^{2+}$ abundance relative to that of the sulfur ions, between
January 2000 and January 2001.  It also appears that the abundance of
Cl$^{+}$ relative to Cl$^{2+}$ increased over this period.
This variability may indicate a primarily volcanic source of Cl in
Io's atmosphere \citep{Lellouch:2003,Moses:2002}.

From the detection of \ion{C}{3} $\lambda$977 discussed in
Section~\ref{other}, we find that [C$^{2+}$]/[S$^{2+}$] $= 3.7 \times
10^{-4}$.  However, this should be regarded as an upper limit to the
abundance of Iogenic carbon in that these ions could have a solar wind
origin \citep{Krimigis:1983}.  From the \ion{C}{2} and \ion{Si}{3}
upper limits, 3$\sigma$ relative column density upper limits are
[C$^{+}$]/[S$^{2+}$] $\leq 2.5 \times 10^{-3}$ and
[Si$^{2+}$]/[S$^{2+}$] $\leq 4.1 \times 10^{-3}$.  From
\citeauthor{Grevesse:1998} the cosmic abundance ratios relative to S
are [C]/[S] = 15.5 and [Si]/[S] = 1.66, so assuming that C$^{2+}$ and
Si$^{2+}$ are also the dominant ion stages, both carbon and silicon are
severely depleted in the torus relative to solar abundances.  For
carbon the depletion is greater than four orders of magnitude and is
likely due to the outgassing of volatiles from Io from continuous
volcanic activity.  For phosphorus an upper limit abundance estimate is
not possible due to the lack of published collision strengths.

\section{CONCLUSION}

The {\it Far Ultraviolet Spectroscopic Explorer} has recorded the
spectrum of the Io plasma torus in the range 905--1187 \AA\ at spectral
resolutions of 0.26 and 0.085~\AA, the highest to date.  Almost all of
the emissions observed arise from ions of sulfur and the details of the
spectra can be matched well by isothermal models using the CHIANTI
database of atomic parameters and multi-level excitation models.  
These models also permit constraints to be set on the electron density
and temperature in the region of the torus where the ultraviolet
emissions are excited.  In addition to sulfur, the resonance multiplets
of \ion{Cl}{2} and \ion{Cl}{3} are clearly detected and we derive an
overall chlorine ion abundance in the torus of slightly less than 1\%,
a decrease relative to that derived from observations made a year
earlier \citep{Feldman:2001a}. \ion{C}{3} $\lambda$977 is detected and
a number of features near 990 \AA\ remain unidentified.  
The source of the carbon ions is uncertain and may be the result of
charge transfer of solar wind ions in the Jovian magnetosphere.  There
is no evidence for ions of nitrogen, silicon, or phosphorus, all which
which have strong lines in the \fuse\ spectral range.  From spectra
obtained at 0.085 \AA\ resolution we are able to measure the linewidths
of the strongest far ultraviolet emissions.  These are found to be
consistent with an ion temperature of 60--70 eV for each of the
observed sulfur ions.

\acknowledgments

We thank Peter Young for valuable discussions and insights into
CHIANTI and Andrew Steffl for making available preprints of the {\it
Cassini} results.  We thank the \fuse\ ground system personnel,
particularly B.\ Roberts, T.\ Ake, A.\ Berman, B.\ Gawne, and
J.\ Andersen, for their efforts in planning and executing these
difficult moving target observations.  This work is based on data
obtained for the Guaranteed Time Team by the NASA-CNES-CSA
\fuse\ mission operated by the Johns Hopkins University.  Financial
support was provided by NASA contract NAS5-32985 and by grant
NAG5-12051 to DFS.  CHIANTI is a collaborative project involving NRL
(USA), RAL (UK), and the Universities of Florence (Italy) and Cambridge
(UK).

%\clearpage
%\bibliography{torus_refs}

\clearpage

\begin{table}
\begin{center}
\caption{Ion Lines Identified in the {\it FUSE} Spectrum of the West Ansa,
14 January 2001  \label{tab:sii}}
\medskip
\begin{tabular}[c]{@{}|lcccccc|@{}}
\tableline
 Ion & Wavelength\tablenotemark{a} & Configuration & Designation & Configuration & Designation &  Brightness\tablenotemark{b} \\
 & (\AA) & ($\ell$) & ($\ell$) & ($u$) & ($u$) &  (rayleighs) \\
\tableline 
\tableline 
\ion{S}{2} & 906.508 & $3s^23p^3$ & $^4$S$_{3/2}$ & $3s^23p^2(^3{\rm P})3d$ & $^4$F$_{5/2}$   & $blend$  \\
\ion{S}{2} & 906.876 & $3s^23p^3$ & $^4$S$_{3/2}$ & $3s^23p^2(^3{\rm P})4s$ & $^4$P$_{5/2}$   & $12.5 \pm 0.30$\tablenotemark{c}   \\
\ion{S}{2} & 910.485 & $3s^23p^3$ & $^4$S$_{3/2}$ & $3s^23p^2(^3{\rm P})4s$ & $^4$P$_{3/2}$ &   $5.93 \pm 0.21$ \\
\ion{S}{2} & 912.736 & $3s^23p^3$ & $^4$S$_{3/2}$ & $3s^23p^2(^3{\rm P})4s$ & $^4$P$_{1/2}$ &   $3.12 \pm 0.16$ \\
\ion{S}{2} & 915.388 & $3s^23p^3$ & $^2$P$_{3/2}$ & $3s^23p^2(^3{\rm P})3d$ & $^2$D$_{5/2}$ &   $0.51 \pm 0.07$ \\
\ion{S}{2} & 918.813 & $3s^23p^3$ & $^2$P$_{1/2}$ & $3s^23p^2(^3{\rm P})3d$ & $^2$D$_{3/2}$ &   $blend$  \\
\ion{S}{2} & 919.208 & $3s^23p^3$ & $^2$P$_{3/2}$ & $3s^23p^2(^3{\rm P})3d$ & $^2$D$_{3/2}$ &   $0.34 \pm 0.04$ \\
\ion{S}{2} & 937.420 & $3s^23p^3$ & $^2$D$_{3/2}$ & $3s^23p^2(^1{\rm D})4s$ & $^2$D$_{3/2}$ &   $blend$  \\
\ion{S}{2} & 937.688 & $3s^23p^3$ & $^2$D$_{5/2}$ & $3s^23p^2(^1{\rm D})4s$ & $^2$D$_{5/2}$ &   $2.77 \pm 0.10$ \\
\ion{S}{2} & 996.007 &  $3s^23p^3$ & $^2$D$_{5/2}$ & $3s^23p^2(^3{\rm P})3d$ & $^2$F$_{7/2}$ &  $4.44 \pm 0.21$ \\
\ion{S}{2} & 1000.486 &  $3s^23p^3$ & $^2$D$_{3/2}$ & $3s^23p^2(^3{\rm P})3d$ & $^2$F$_{5/2}$ &  $3.20 \pm 0.13$ \\
\ion{S}{2} & 1006.091 &  $3s^23p^3$ & $^2$D$_{5/2}$ & $3s^23p^2(^3{\rm P})3d$ & $^4$D$_{7/2}$ & $blend$ \\
\ion{S}{2} & 1006.258 &  $3s^23p^3$ & $^2$D$_{3/2}$ & $3s^23p^2(^3{\rm P})3d$ & $^4$D$_{5/2}$ & $blend$ \\
\ion{S}{2} & 1006.580 &  $3s^23p^3$ & $^2$D$_{5/2}$ & $3s^23p^2(^3{\rm P})3d$ & $^4$D$_{5/2}$ & $10.1 \pm 0.16$ \\
\ion{S}{2} & 1006.889 &  $3s^23p^3$ & $^2$D$_{5/2}$ & $3s^23p^2(^3{\rm P})3d$ & $^4$D$_{3/2}$ & $blend$ \\
\ion{S}{2} & 1006.954 &  $3s^23p^3$ & $^2$D$_{3/2}$ & $3s^23p^2(^3{\rm P})3d$ & $^4$D$_{1/2}$ & $1.66 \pm 0.07$ \\
\ion{S}{2} & 1014.437 &  $3s^23p^3$ & $^2$D$_{5/2}$ & $3s^23p^2(^3{\rm P})4s$ & $^2$P$_{3/2}$ & $3.30 \pm 0.09$ \\
\ion{S}{2} & 1019.528 &  $3s^23p^3$ & $^2$D$_{3/2}$ & $3s^23p^2(^3{\rm P})4s$ & $^2$P$_{1/2}$ & $1.74 \pm 0.06$ \\
\ion{S}{2} & 1042.950 &  $3s^23p^3$ & $^2$D$_{5/2}$ & $3s^23p^2(^3{\rm P})3d$ & $^4$F$_{9/2}$ & $2.09 \pm 0.09$ \\
\ion{S}{2} & 1045.763 &  $3s^23p^3$ & $^2$D$_{5/2}$ & $3s^23p^2(^3{\rm P})3d$ & $^4$F$_{7/2}$ & $7.84 \pm 0.11$ \\
\ion{S}{2} & 1047.554 &  $3s^23p^3$ & $^2$D$_{3/2}$ & $3s^23p^2(^3{\rm P})3d$ & $^4$F$_{5/2}$  & $blend$ \\
\ion{S}{2} & 1047.903 &  $3s^23p^3$ & $^2$D$_{5/2}$ & $3s^23p^2(^3{\rm P})3d$ & $^4$F$_{5/2}$  & $2.08 \pm 0.07$ \\
\ion{S}{2} & 1049.053 &  $3s^23p^3$ & $^2$D$_{3/2}$ & $3s^23p^2(^3{\rm P})3d$ & $^4$F$_{3/2}$ &  $3.12 \pm 0.08$ \\
\ion{S}{2} & 1096.596 &  $3s^23p^3$ & $^2$D$_{3/2}$ & $3s^23p^2(^3{\rm P})3d$ & $^2$P$_{1/2}$ &  $5.40 \pm 0.10$ \\
\ion{S}{2} & 1102.362 &  $3s^23p^3$ & $^2$D$_{5/2}$ & $3s^23p^2(^3{\rm P})3d$ & $^2$P$_{3/2}$ &  $11.3 \pm 0.13$ \\
\ion{S}{2} & 1124.395 & $3s^23p^3$ & $^2$P$_{1/2}$ & $3s^23p^2(^3{\rm P})4s$ & $^2$P$_{3/2}$ &  $0.21 \pm 0.02$ \\
\ion{S}{2} & 1124.986 & $3s^23p^3$ & $^2$P$_{3/2}$ & $3s^23p^2(^3{\rm P})4s$ & $^2$P$_{3/2}$ &   $1.14 \pm 0.05$ \\
\ion{S}{2} & 1131.059 & $3s^23p^3$ & $^2$P$_{1/2}$ & $3s^23p^2(^3{\rm P})4s$ & $^2$P$_{1/2}$ &  $0.39 \pm 0.03$ \\
\ion{S}{2} & 1131.657 & $3s^23p^3$ & $^2$P$_{3/2}$ & $3s^23p^2(^3{\rm P})4s$ & $^2$P$_{1/2}$ &  $0.20 \pm 0.02$ \\
\ion{S}{2} & 1166.291 & $3s^23p^3$ & $^2$P$_{3/2}$ & $3s^23p^2(^3{\rm P})3d$ & $^4$F$_{5/2}$ &  $0.23 \pm 0.03$ \\
\ion{S}{2} & 1167.512 & $3s^23p^3$ & $^2$P$_{1/2}$ & $3s^23p^2(^3{\rm P})3d$ & $^4$F$_{3/2}$ &   $0.68 \pm 0.05$ \\
\ion{S}{2} & 1168.150 & $3s^23p^3$ & $^2$P$_{3/2}$ & $3s^23p^2(^3{\rm P})3d$ & $^4$F$_{3/2}$ &   $0.85 \pm 0.05$ \\
\ion{S}{2} & 1172.882 & $3s^23p^3$ & $^2$P$_{3/2}$ & $3s^23p^2(^3{\rm P})4s$ & $^4$P$_{3/2}$ &  $0.30 \pm 0.04$ \\
\tableline
\end{tabular}
\vspace*{-0.5cm}
\tablenotetext{a}{Laboratory wavelength.}
\tablenotetext{b}{Average brightness in $30'' \times 30''$ aperture.}
\tablenotetext{c}{Standard deviation in observed counts.}
\end{center}
\end{table}

\begin{table}
\begin{center}
\caption{Ion Lines Identified in the {\it FUSE} Spectrum of the West Ansa 
-- continued \label{tab:siii}}
\medskip
\begin{tabular}[c]{@{}|lcccccc|@{}}
\tableline
 Ion & Wavelength\tablenotemark{a} & Configuration & Designation & Configuration & Designation &  Brightness\tablenotemark{b} \\
 & (\AA) & ($\ell$) & ($\ell$) & ($u$) & ($u$) &  (rayleighs) \\
\tableline 
\tableline 
\ion{S}{3} & 1012.492 &  $3s^23p^2$ & $^3$P$_{0}$ & $3s3p^3$ & $^3$P$_{1}$ & $9.88 \pm 0.13$\tablenotemark{c}    \\
\ion{S}{3} & 1015.496 &  $3s^23p^2$ & $^3$P$_{1}$ & $3s3p^3$ & $^3$P$_{0}$  & $blend$ \\
\ion{S}{3} & 1015.561 &  $3s^23p^2$ & $^3$P$_{1}$ & $3s3p^3$ & $^3$P$_{1}$  & $blend$ \\
\ion{S}{3} & 1015.775 &  $3s^23p^2$ & $^3$P$_{1}$ & $3s3p^3$ & $^3$P$_{2}$  & $23.2 \pm 0.16$ \\
\ion{S}{3} & 1016.85 & \multicolumn{4}{c}{{\it configuration not identified in \citet{Kelly:1987}}} &    $0.97 \pm 0.05$ \\
\ion{S}{3} & 1021.105 &  $3s^23p^2$ & $^3$P$_{2}$ & $3s3p^3$ & $^3$P$_{1}$  & $blend$ \\
\ion{S}{3} & 1021.321 &  $3s^23p^2$ & $^3$P$_{2}$ & $3s3p^3$ & $^3$P$_{2}$  & $38.3 \pm 0.16$ \\
\ion{S}{3} & 1077.171 &  $3s^23p^2$ & $^1$D$_{2}$ & $3s^23p3d$ & $^1$D$_{2}$ & $30.7 \pm 0.20$ \\
\ion{S}{3} & 1121.755 & $3s3p^3$ & $^3$D$_{2}$ & $3s^23p4p$ & $^3$P$_{2}$ &    $0.11 \pm 0.02$ \\
\ion{S}{3} & 1122.418 & $3s3p^3$ & $^3$D$_{3}$ & $3s^23p4p$ & $^3$P$_{2}$ & $0.54 \pm 0.04$ \\
\ion{S}{3} & 1126.536 & $3s3p^3$ & $^3$D$_{1}$ & $3s^23p4p$ & $^3$P$_{1}$ & $blend$   \\
\ion{S}{3} & 1126.886 & $3s3p^3$ & $^3$D$_{2}$ & $3s^23p4p$ & $^3$P$_{1}$  & $1.01 \pm 0.05$ \\
\ion{S}{3} & 1128.500 & $3s3p^3$ & $^3$D$_{1}$ & $3s^23p4p$ & $^3$P$_{0}$ & $1.08 \pm 0.05$ \\
\tableline 
\tableline
\ion{S}{4} & 1062.664 &  $3s^23p$ & $^2$P$_{1/2}$ & $3s3p^2$ & $^2$D$_{3/2}$ & $21.2 \pm 0.16$ \\
\ion{S}{4} & 1072.974 &  $3s^23p$ & $^2$P$_{3/2}$ & $3s3p^2$ & $^2$D$_{5/2}$ & $16.9 \pm 0.16$ \\
\ion{S}{4} & 1073.518 &  $3s^23p$ & $^2$P$_{3/2}$ & $3s3p^2$ & $^2$D$_{3/2}$ & $3.48 \pm 0.09$ \\
\ion{S}{4} & 1088.83 & \multicolumn{4}{c}{{\it identified by \citet{Kelly:1987}, configuration uncertain}} &    $0.90 \pm 0.05$ \\
\ion{S}{4} & 1098.929 &  $3s3p^2$ & $^2$D$_{5/2}$ & $3p^3$ & $^2$D$_{5/2}$  &  $0.45 \pm 0.04$ \\
\ion{S}{4} & 1099.480 &  $3s3p^2$ & $^2$D$_{3/2}$ & $3p^3$ & $^2$D$_{3/2}$   & $0.40 \pm 0.03$ \\
\tableline 
\tableline
\ion{Cl}{2} & 1063.831 & $3s^23p^4$ & $^3$P$_{2}$ & $3s3p^5$ & $^3$P$_{1}$ & $0.17 \pm 0.02$ \\
\ion{Cl}{2} & 1071.036 & $3s^23p^4$ & $^3$P$_{2}$ & $3s3p^5$ & $^3$P$_{2}$ & $0.46 \pm 0.04$ \\
\ion{Cl}{2} & 1079.080 & $3s^23p^4$ & $^3$P$_{1}$ & $3s3p^5$ & $^3$P$_{2}$ & $0.19 \pm 0.03$ \\
\tableline 
\tableline
\ion{Cl}{3} & 1005.28 & $3s^23p^3$ & $^4$S$_{3/2}$ & $3s3p^4$ & $^4$P$_{1/2}$ & $0.36 \pm 0.04$  \\
\ion{Cl}{3} & 1008.78 & $3s^23p^3$ & $^4$S$_{3/2}$ & $3s3p^4$ & $^4$P$_{3/2}$ & $0.73 \pm 0.05$ \\
\ion{Cl}{3} & 1015.02 & $3s^23p^3$ & $^4$S$_{3/2}$ & $3s3p^4$ & $^4$P$_{5/2}$ & $1.39 \pm 0.07$ \\
\tableline
\end{tabular}
\vspace*{-0.5cm}
\tablenotetext{a}{Laboratory wavelength.}
\tablenotetext{b}{Average brightness in $30'' \times 30''$ aperture.}
\tablenotetext{c}{Standard deviation in observed counts.}
\end{center}
\end{table}

\clearpage 
\begin{center}{\bf FIGURE CAPTIONS}\end{center}

\figcaption{\fuse\ spectrum of the west ansa of the Io torus obtained
on 2001 January 14 showing the regions containing the \ion{Cl}{3} (top)
and  \ion{Cl}{2} (bottom) multiplets.  The total exposure time was
9740~s.  The gray curve is the data scaled by a factor of 10.
\label{spec1}}

\figcaption{Same as Figure~\ref{spec1} for the spectral region from 905
to 940 \AA\ showing the \ion{S}{2} second resonance triplet.
\label{spec2}}

\figcaption{Same as Figure~\ref{spec1} for the spectral region from
1085 to 1135 \AA\ showing the highly excited \ion{S}{3} and \ion{S}{4}
multiplets.
\label{spec3}}

\figcaption{Same as Figure~\ref{spec1} for the spectral region from 970
to 1000 \AA\ showing spectra from three independent channels.  Unidentified
features that appear in at least two channels are denoted by ``?''.
\label{spec4}}

\figcaption{\fuse\ spectrum from the MDRS aperture obtained on 2001 October
19 and 21.  The same spectral region as the top panel of Figure~\ref{spec1} is
shown from two separate channels.  The LiF2b channel has about one-half
the effective area as the LiF1a channel.  The total integration time
was 31,333~s.  The LiF1a spectrum also includes several \htwo\ emissions
from the Jovian disk.
\label{spec_mdrs}}

\figcaption{Calculated line ratios as a function of \te\ for \ion{S}{3}
$\lambda\lambda$1128.5, 1126.8, and 1122.4 (a, b, and c, respectively)
relative to $\lambda$1012.5 and \ion{S}{4} $\lambda\lambda$1099.5 and
1098.9 (d and e, respectively), relative to $\lambda$1062.7.
A plasma with spatially homogeneous, constant electron temperature defined 
by a single Maxwellian distribution and \nelec\ =~2000 \cc\ was assumed.
The CHIANTI v.4 software was used.  The observed ratios are indicated
by crosses (LiF1b) and diamonds (LiF2a).  The error bars are 1-$\sigma$
in the observed counts.  \label{tedep}}

\figcaption{Calculated line ratios as a function of \nelec\ for
\ion{S}{2} and \ion{S}{3} lines indicated in the figure for two values
of \te, 50,000 K (solid line) and 80,000 K (dashed line).  The CHIANTI
v.4 software was used.  The observed ratios are from the LiF1a
channel.  The error bars are 1-$\sigma$ in the observed counts.
\label{nedep}}

\setcounter{figure}{0}
\begin{figure*}
\begin{center}
\epsscale{0.85}
\plotone{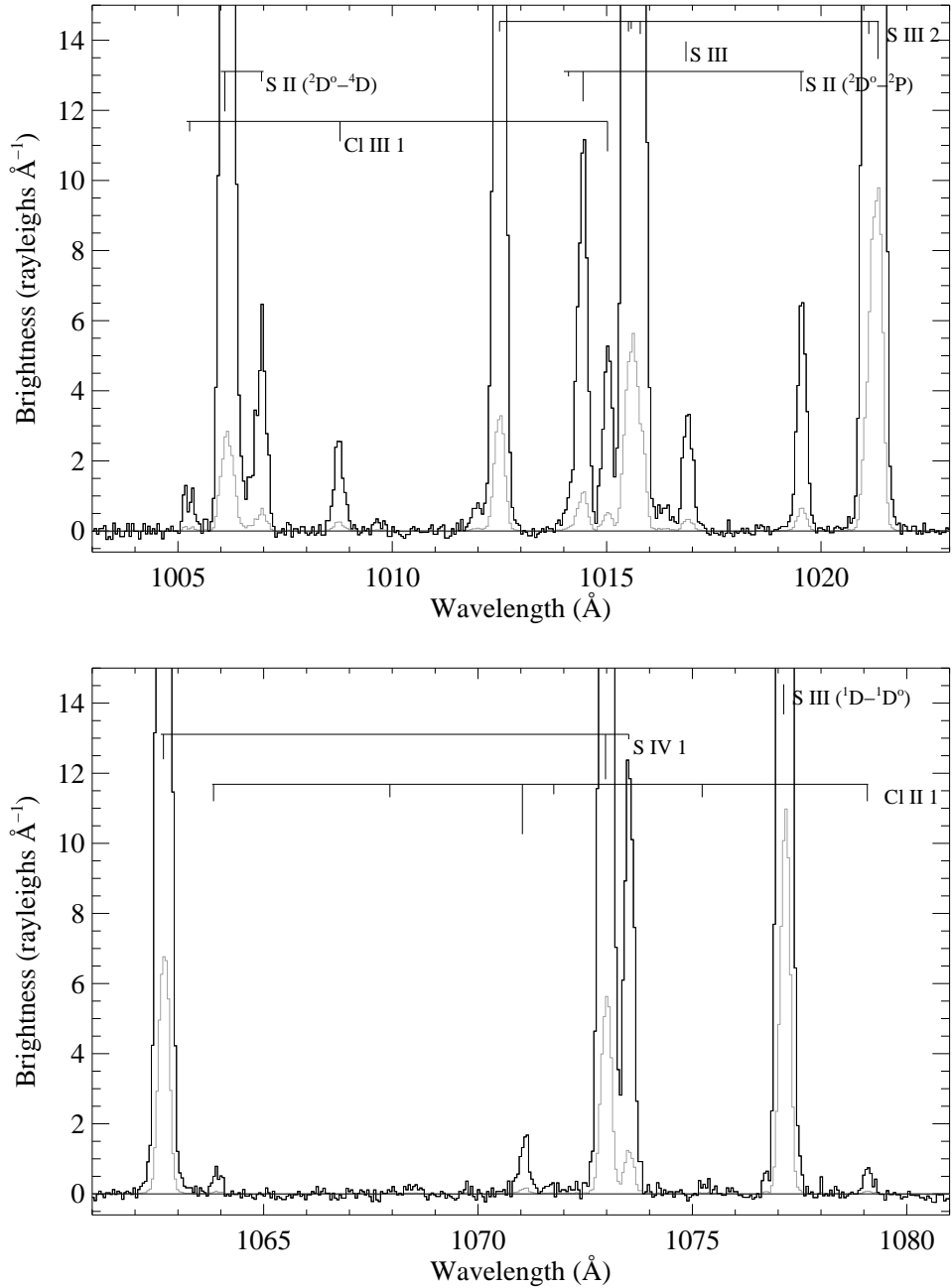}
\vspace{0.1in}
\caption{\fuse\ spectrum of the west ansa of the Io torus obtained
on 2001 January 14 showing the regions containing the \ion{Cl}{3} (top)
and  \ion{Cl}{2} (bottom) multiplets.  The total exposure time was
9740~s.  The gray curve is the data scaled by a factor of 10.}
\end{center}
\end{figure*}

\begin{figure*}
\begin{center}
\epsscale{0.9}
\plotone{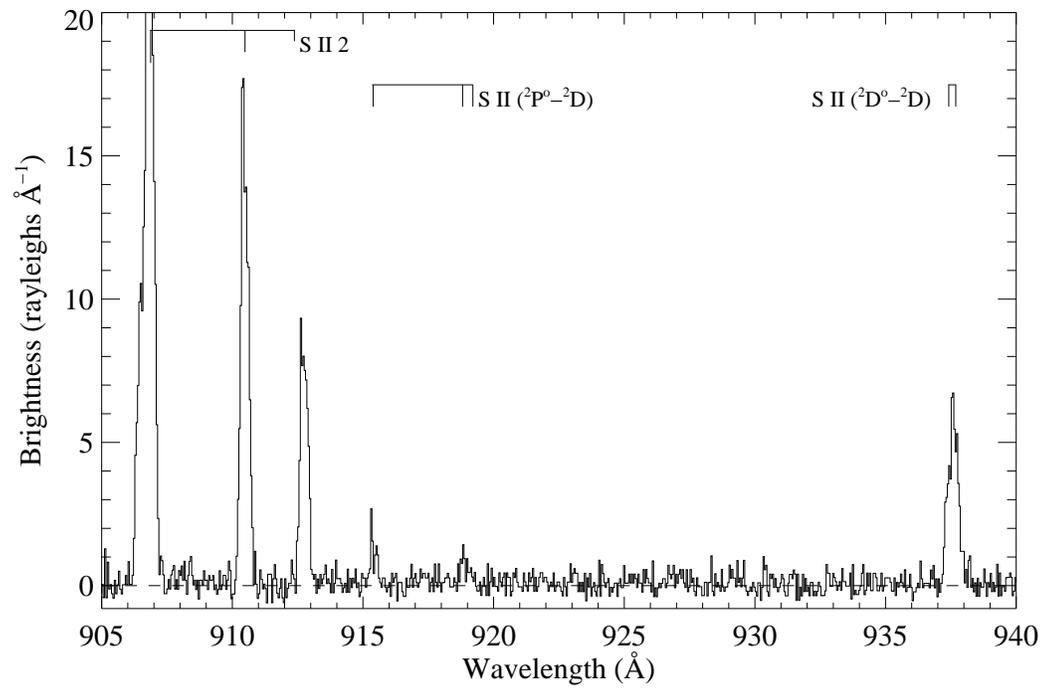}
\vspace{0.1in}
\caption{Same as Figure~\ref{spec1} for the spectral region from 905
to 940 \AA\ showing the \ion{S}{2} second resonance triplet.}
\end{center}
\end{figure*}

\begin{figure*}
\begin{center}
\epsscale{0.9}
\plotone{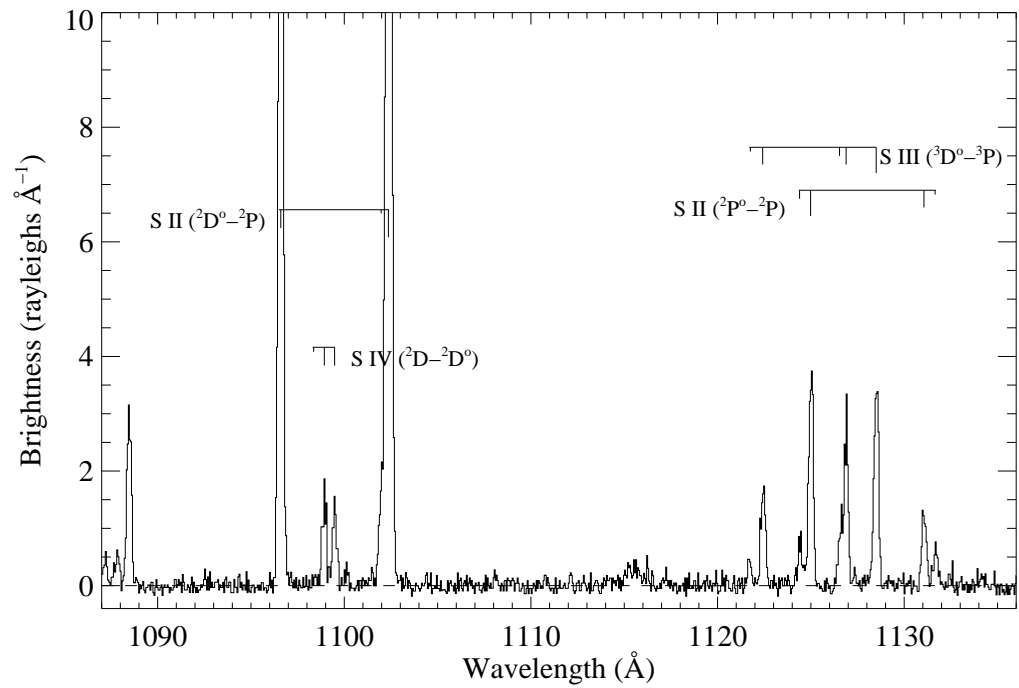}
\vspace{0.1in}
\caption{Same as Figure~\ref{spec1} for the spectral region from
1085 to 1135 \AA\ showing the highly excited \ion{S}{3} and \ion{S}{4}
multiplets.}
\end{center}
\end{figure*}

\begin{figure*}
\begin{center}
\epsscale{0.9}
\plotone{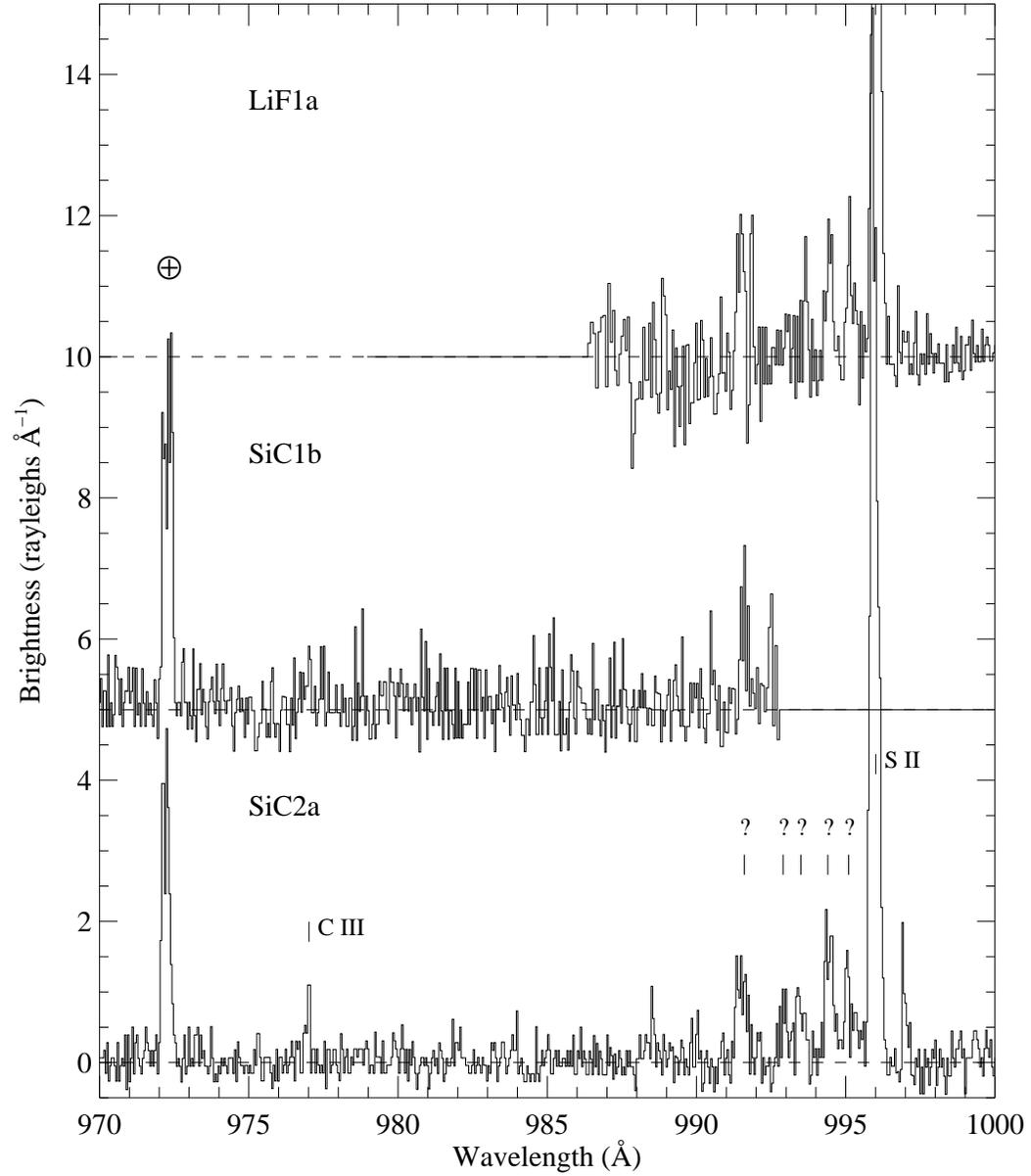}
\vspace{0.1in}
\caption{Same as Figure~\ref{spec1} for the spectral region from 970
to 1000 \AA\ showing spectra from three independent channels.  Unidentified
features that appear in at least two channels are denoted ?.}
\end{center}
\end{figure*}

\begin{figure*}
\begin{center}
\epsscale{0.9}
\plotone{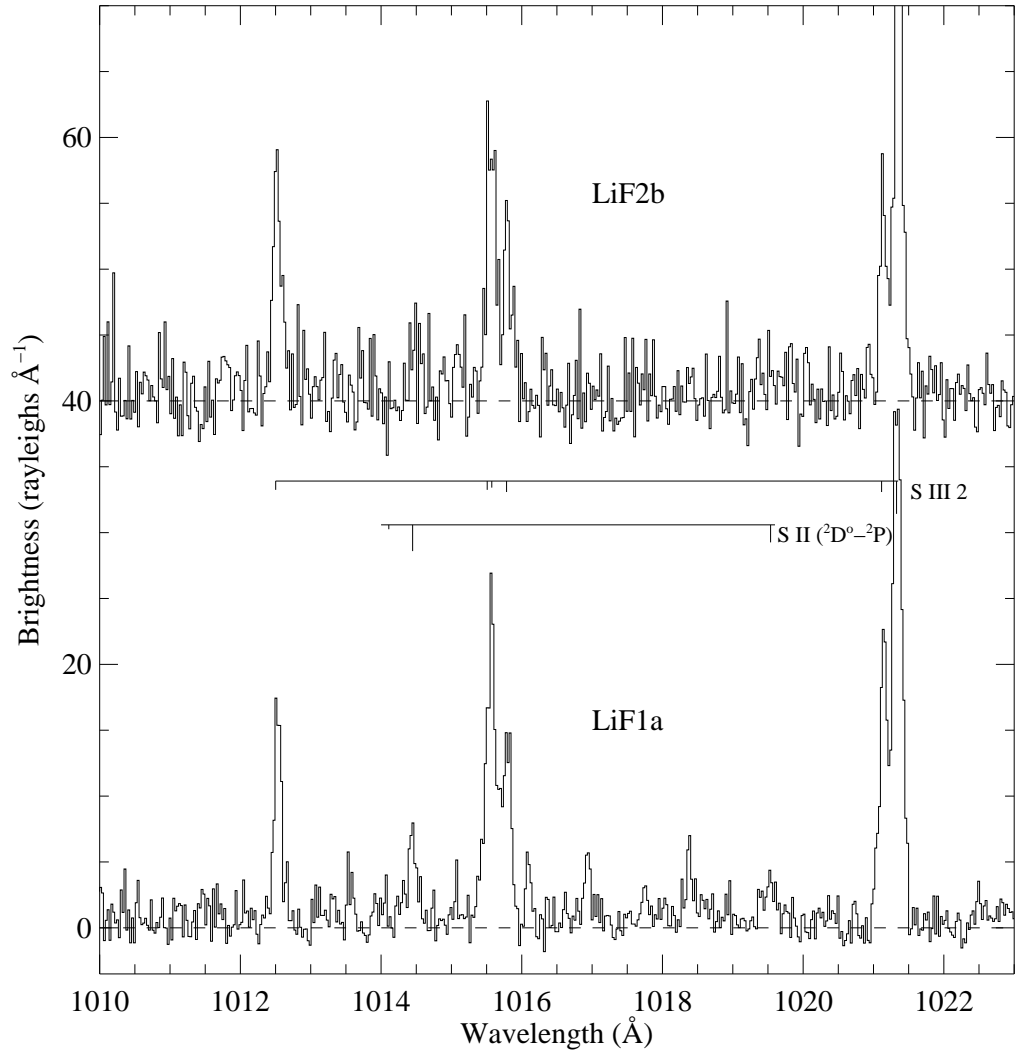}
\vspace{0.1in}
\caption{\fuse\ spectrum from the MDRS aperture obtained on 2001 October
19 and 21.  The same spectral region as the top panel of Figure~\ref{spec1} is
shown from two separate channels.  The LiF2b channel has about one-half
the effective area as the LiF1a channel.  The total integration time
was 31,333~s.  The LiF1a spectrum also includes several \htwo\ emissions
from the Jovian disk.}
\end{center}
\end{figure*}

\begin{figure*}
\begin{center}
\epsscale{0.9}
\plotone{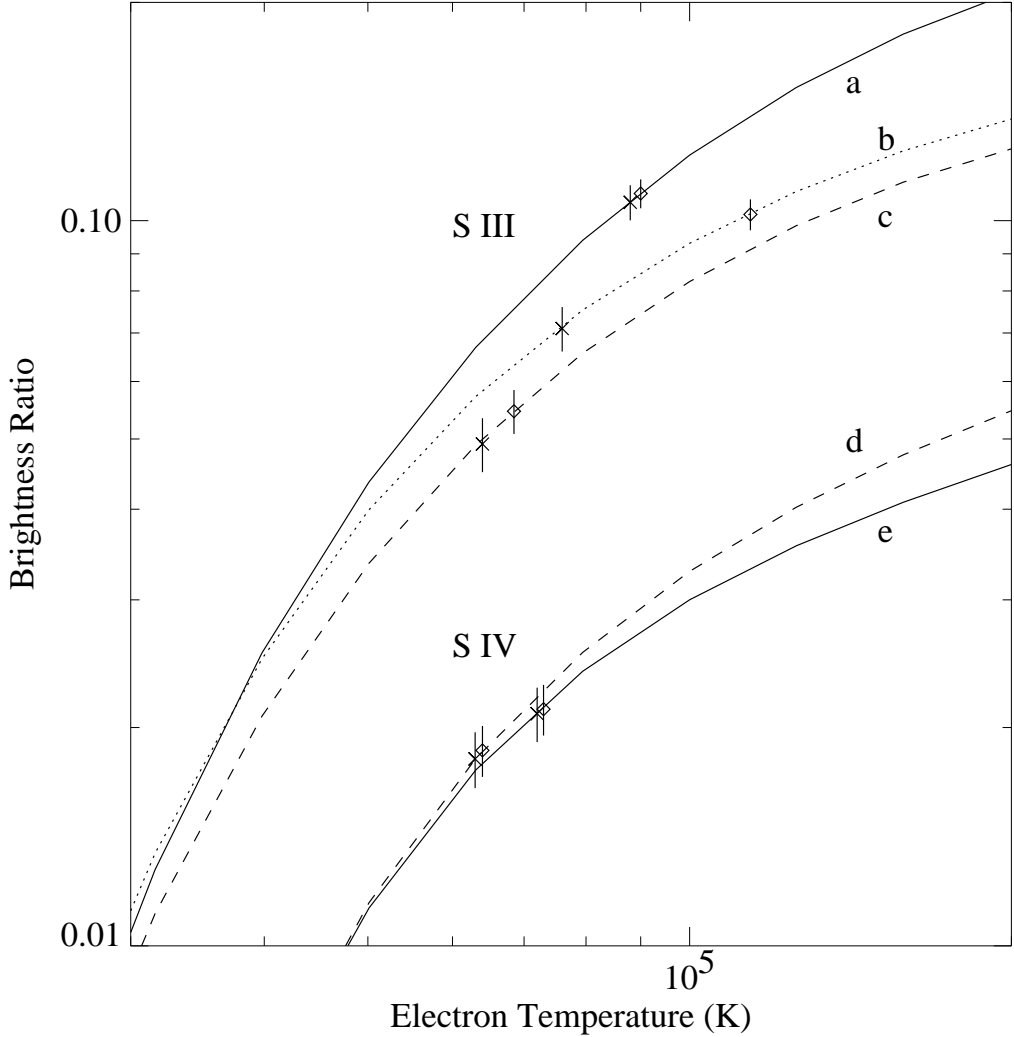}
\vspace{0.1in}
\caption{Calculated line ratios as a function of \te\ for \ion{S}{3}
$\lambda\lambda$1128.5, 1126.8, and 1122.4 (a, b, and c, respectively)
relative to $\lambda$1012.5 and \ion{S}{4} $\lambda\lambda$1099.5 and
1098.9 (d and e, respectively), relative to $\lambda$1062.7.
A plasma with spatially homogeneous, constant electron temperature defined 
by a single Maxwellian distribution and \nelec\ =~2000 \cc\ was assumed.
The CHIANTI v.4 software was used.  The observed ratios are indicated
by crosses (LiF1b) and diamonds (LiF2a).  The error bars are 1-$\sigma$
in the observed counts.}
\end{center}
\end{figure*}

\begin{figure*}
\begin{center}
\epsscale{0.9}
\plotone{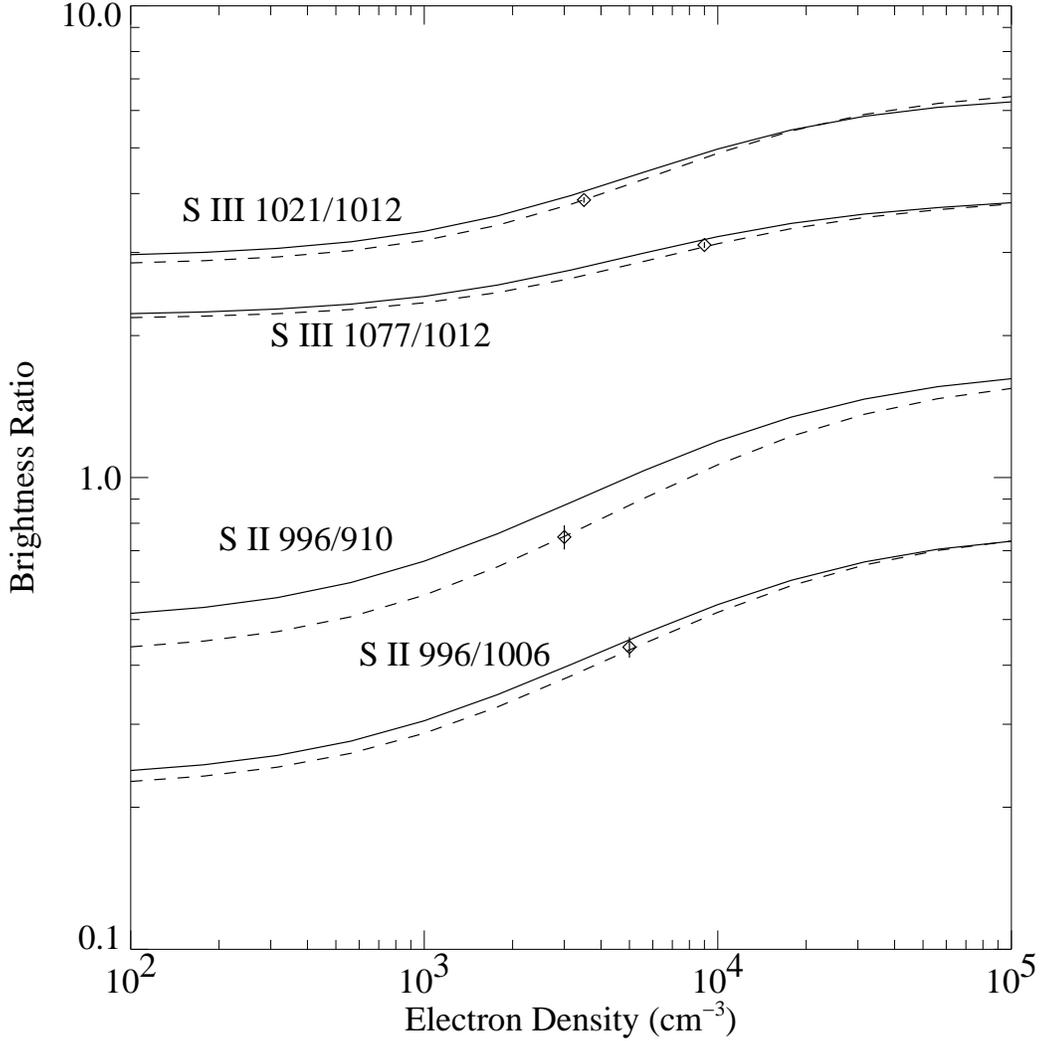}
\vspace{0.1in}
\caption{Calculated line ratios as a function of \nelec\ for
\ion{S}{2} and \ion{S}{3} lines indicated in the figure for two values
of \te, 50,000 K (solid line) and 80,000 K (dashed line).  The CHIANTI
v.4 software was used.  The observed ratios are from the LiF1a
channel.  The error bars are 1-$\sigma$ in the observed counts.}
\end{center}
\end{figure*}

\end{document}